\documentclass[11pt]{article}
\usepackage{moriond,epsfig}

\def\be{\begin{equation}}
\def\ee{\end{equation}}
\def\bea{\begin{eqnarray}}
\def\eea{\end{eqnarray}}

\begin{document}
\vspace*{4cm}

\title{LATEST JET RESULTS FROM TEVATRON}

\author{A. MESSINA}

\address{(on behalf of the CDF and D\O~collaborations)\\
INFN sezione di Roma, Piazzale Aldo Moro 2,
00185 Roma, Italy}

\maketitle
\abstracts{
This contribution reports preliminary jet results in $p\bar{p}$ collisions at
$\sqrt{s}=1.96$ TeV from the CDF and D\O~ experiments.
The jet inclusive cross section, measured using both the Midpoint and the $K_T$
jet clustering algorithm, is compared to next-to-leading order QCD prediction in
different rapidity regions.
The b-jet inclusive cross section measured exploiting the long lifetime and large
mass of B hadrons is presented and compared to QCD prediction. A complementary measurement,
using the large branching fraction of B hadrons into muons, is also described.
The measurement of two-particle momentum correlation in jets is presented and compared
to predictions.
}

The QCD program at the Tevatron hadron collider revolves around the jet physics.
Jets are collimated sprays of hadrons generated by the fragmentation of partons
originating from the hard scattering. In hadron collisions, jets receive  soft
contributions  from the initial state radiation and the multiple
parton interaction between the beam remnants (underlying event (UE)).
Jets are experimentally observed by adding the energy measured in each calorimeter cell
 associated to a cluster by a defined jet algorithm. Both the CDF~\cite{ref:cdf} and
the D\O~\cite{ref:d0} collaboration are exploring new and alternative jet algorithms 
with respect to the cone-based used in Run I (JetClu) that is not collinear and infrared safe.  
Jets are now reconstructed with the  $k_T$ algorithm~\cite{ref:kt} or the MidPoint algorithm~\cite{ref:mp}.
The latter is an iterative seed-based cone algorithm but it uses midpoints
between pairs of protojets as additional seeds in order to make the clusterization
procedure infrared safe.
The $k_T$ algorithm merges pairs of nearby protojets in order of increasing relative transverse
momentum, it is infrared and collinear safe to all orders in pQCD.
Regardless of the jet algorithm used, proper comparisons with the theory require corrections
for non-perturbative contributions. These contributions come from the underlying event and the
hadronization processes and become more and more important as $p ^{jet}_T$ decreases.
\begin{figure}
\begin{center}
\includegraphics[height=.29\textheight]{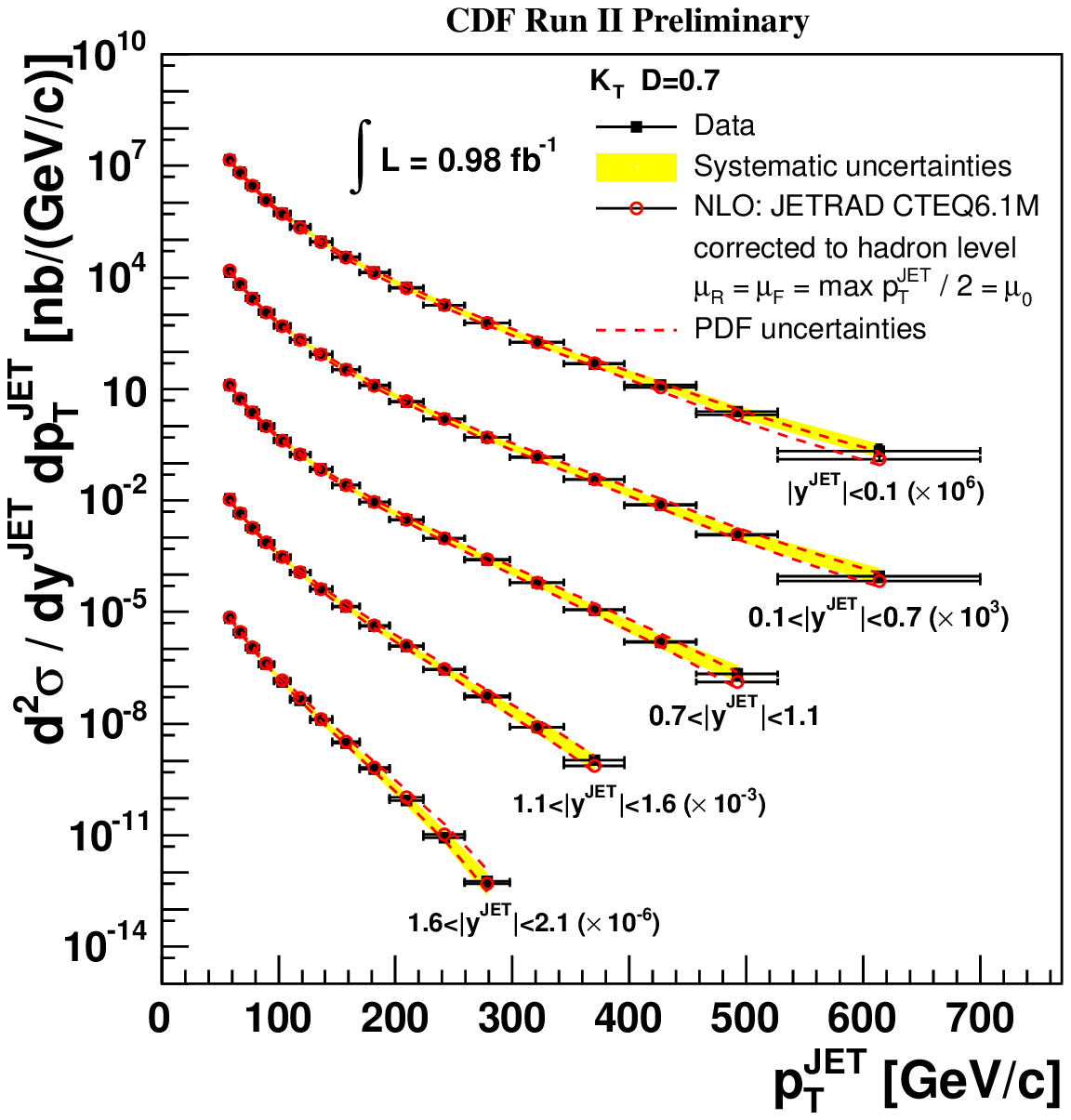}
\includegraphics[height=.31\textheight,width=.31\textheight]{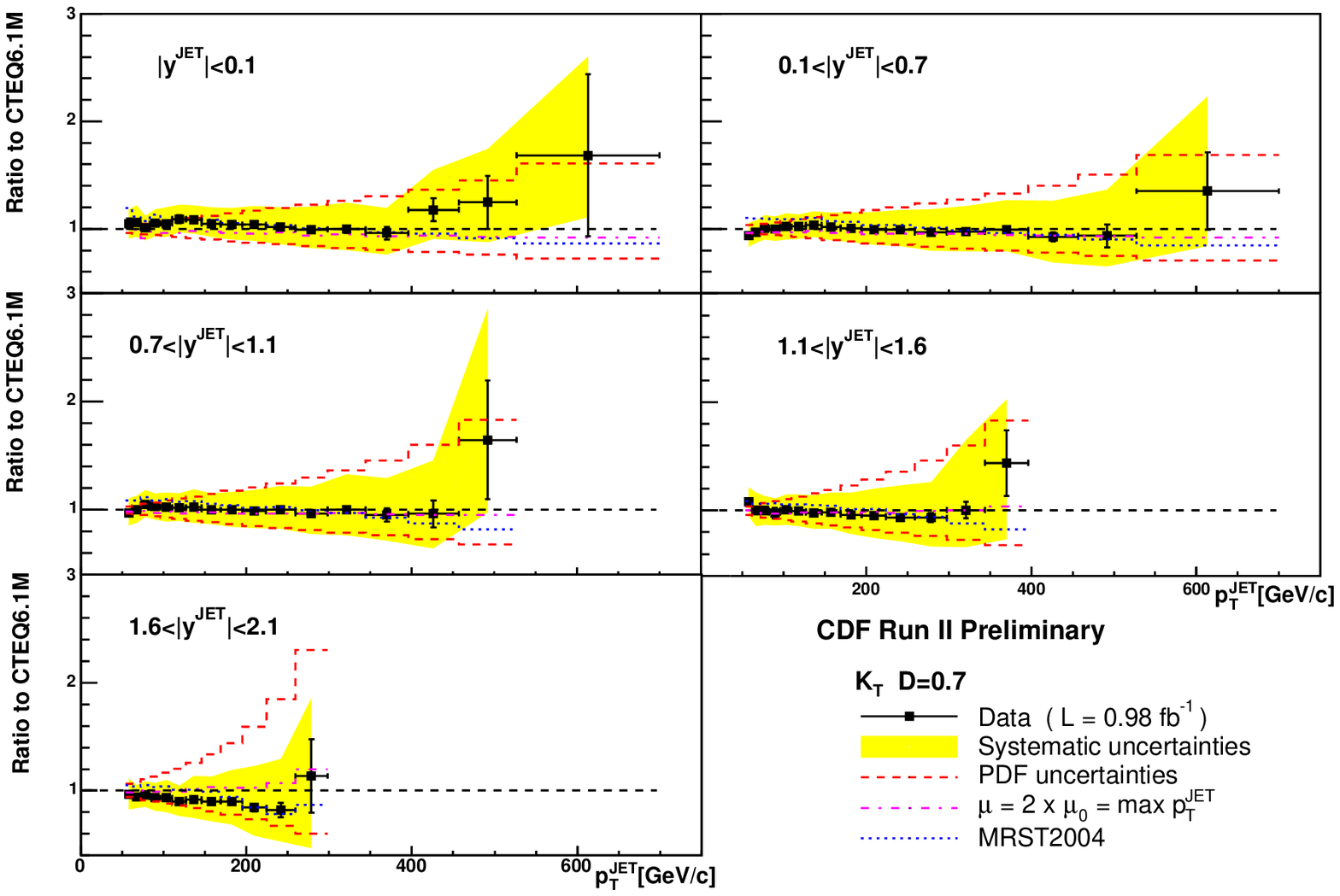}
\caption{Left: inclusive jet cross section measured using the $k_T$ algorithm.
Right: ratio of measured and theoretical cross sections.}
\label{Fig:kt}
\end{center}
\end{figure}
The Underlying event has been studied by CDF in Run I \cite{ref:field} and Run II.
To phenomenologically reproduce the activity of the underlying event with
the PYTHIA~\cite{ref:pythia} Monte Carlo a special set of parameters (Tune A) has been determined.
Tune A has been shown to properly describe the jet shapes~\cite{ref:js} measured in Run II.

To further investigate the fuzzy boundary between domains of perturbative QCD and
non-perturbative hadronization processes, CDF has studied the two-particle momentum
correlation inside jets.
The measurement is based on $\sim400 pb^{-1}$ of dijet data.
The correlation function is introduced in terms of the variable $\xi=log(E_{jet})/P_{hadron}$
and is measured as the ratio of two- and one-particle inclusive momentum distribution functions:
\begin{equation}
C(\xi_1,\xi_2) = \frac{D(\xi_1,\xi_2)}{D(\xi_1)D(\xi_2)};~~ D(\xi_1,\xi_2)=\frac{d^2N}{d\xi_1d\xi_2},~~
D(\xi_1)=\frac{dN}{d\xi_1}
\end{equation}
Only charged particles in a restricted cone with a smaller opening angle of $\theta_c=0.5$
around the jet axis are considered.
The theoretical prediction for the correlation function is~\cite{ref:NLLA}:
\begin{equation}
C(\Delta\xi_1,\Delta\xi_2) =C_0 + C_1(\Delta\xi_1+\Delta\xi_2) +C_2(\Delta\xi_1-\Delta\xi_2)^2;~~
\Delta\xi = \xi-\xi_0
\end{equation}
$\xi_0$ being the position of the peak of the inclusive particle momentum distribution in jets.
The parameters $C_0$, $C_1$ and $C_2$ define the strength of the correlation and depend on a variable
$log(Q/Q_{eff})$ where $Q=E_{jet}\theta_c$ is the jet hardness and $Q_{eff}$
is the parton shower cut-off scale used in theory.
The evolution of the $C_1$ and $C_2$ parameters as a function of jet hardness Q is fitted
with the analytical next-to-leading Log approximation (NLLA) function leaving
$Q_{eff}$ as  a free parameter.
The value of $Q_{eff}$ obtained from the fit
of $C_1$ is $\sim147\pm10(stat)\pm79(syst) MeV$, and from the fit of $C_2$ is
$\sim131\pm12(stat)\pm86(syst) MeV$. The results are found to be in agreement with the
re-summed NLLA calculation.
Correlation clearly survives the hadronization process giving further
support to the hypothesis of local-parton-hadron-duality.
\begin{figure}
\begin{center}
\includegraphics[height=.24\textheight]{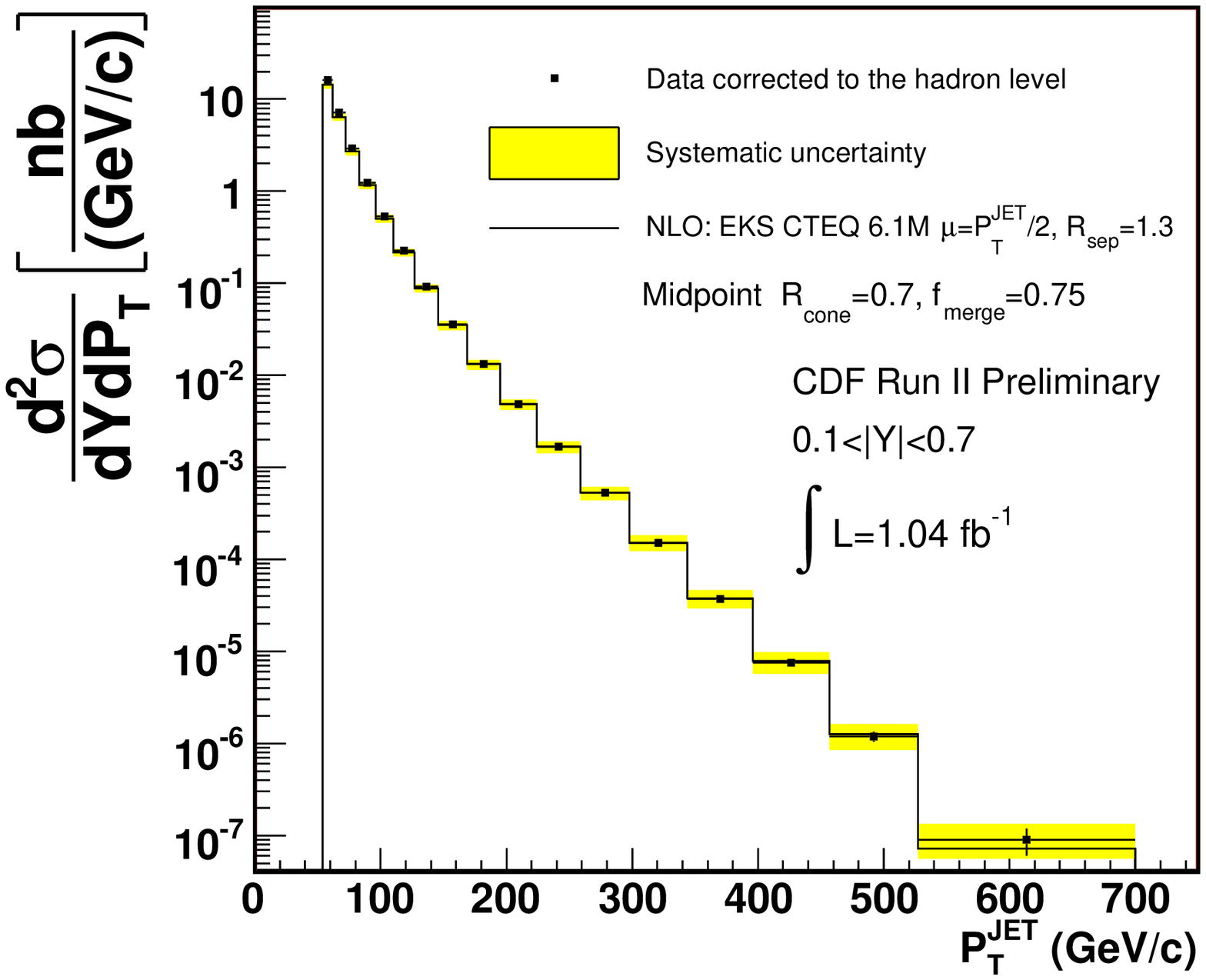}
\includegraphics[height=.24\textheight]{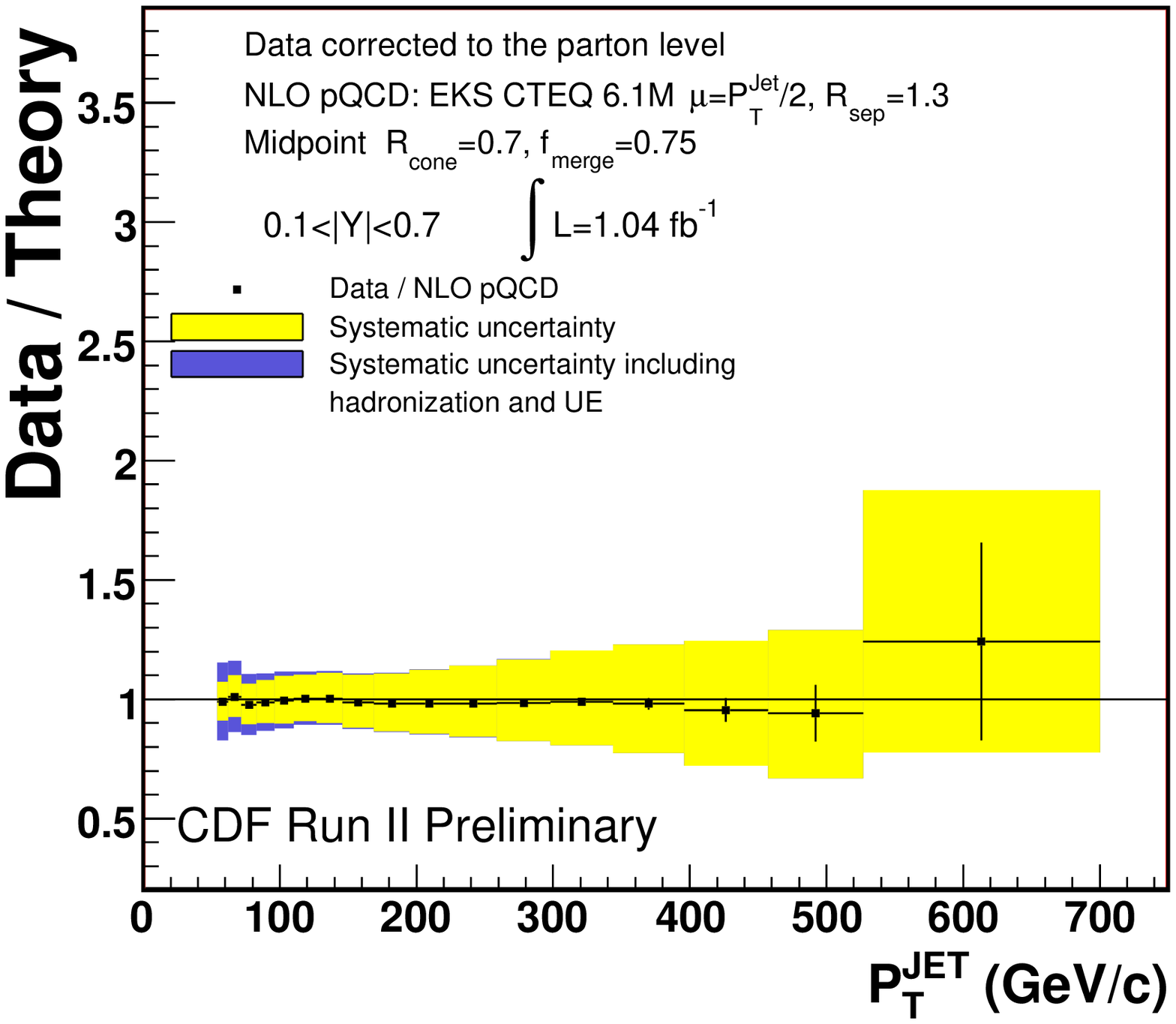}
\caption{Left: inclusive jet cross section measured using the MidPoint algorithm.
Right: ratio of measured and theoretical cross sections.}
\label{Fig:mp}
\end{center}
\end{figure}

The measurement of the inclusive jet production cross section represents a
fundamental test of perturbative QCD predictions over almost nine orders of magnitude.
The high $p^{jet}_T$ tail probes distances down to about $10^{-19}m$  and
is sensitive to new physics. This measurement helps constrain the Parton
Distribution Functions (PDF) at high $x$ and high $Q^2$. In particular at large
rapidities, where no effect from new physics are expected, jet measurements can reduce the uncertainty on 
the gluon density in the proton.
The preliminary results presented here use  $\sim1fb^{-1}$ of data collected at CDF during Run II.
The jet production rate at high $p^{jet}_T$ has significantly increased thanks to
the higher Tevatron center of mass energy, from 1.8 TeV in Run I to 1.96 TeV in Run II,
extending the $p^ {jet}_T$ coverage by about 150 GeV/c.
In order to compare with NLO pQCD, the measured cross section has been corrected
for non perturbative effects. The corresponding parton-to-hadron correction is determined with
PYTHIA Tune A as the ratio of the predicted inclusive jet cross sections at the hadron level with UE
and at the parton level without UE.
This correction was also evaluated with HERWIG~\cite{ref:herwig}.
The difference between the two Monte Carlos was considered as the systematic uncertainty on the correction.
Figure \ref{Fig:kt} shows the inclusive jet cross section measured using the $k_T$ algorithm
in 5 different rapidity regions up to $|y|<2.1$.  The measured cross sections are compared to NLO pQCD
obtained with JETRAD~\cite{ref:jetrad} using CTEQ6.1M~\cite{ref:pdf} PDF
and setting the renormalization and factorization scales
to $max(p ^{jet}_T/2)$. Similarly, figure \ref{Fig:mp} shows the comparison between data and theory using
the MidPoint algorithm with a cone radius of 0.7.
In this case the measurement covers $0.1\le|y|\le0.7$, and the NLO pQCD cross
section was obtained with EKS~\cite{ref:eks}.
An additional  6\% normalization uncertainty associated with the luminosity measurement is not
included on both figures. The experimental uncertainties are dominated by the uncertainty on the
absolute jet energy scale which is known at the level of  2\% at low $p^{jet}_T$ and  3\% at
high $p^{jet}_T$~\cite{ref:nim}. The main uncertainty in the pQCD prediction comes from the PDF,
especially from the limited knowledge of the gluon PDF at high $x$. The
uncertainty on the parton-to-hadron correction factor is also important at low $p^{jet}_T$.
For both the $k_T$ and the midpoint algorithms, the measured cross sections are in good agreement
with the predictions.  A jet inclusive cross section measurement has also been performed by the
D\O~ collaboration using $\sim380pb^{-1}$~\cite{ref:mpd0}.
Here jet are clustered with the MidPoint algorithm. In the
2 rapidity regions covered by the measurement  ($|y|\le0.4$ and $0.4\le|y|\le0.8$) there is good agreement
with NLO pQCD.

\begin{figure}
\begin{center}
\includegraphics[height=.20\textheight]{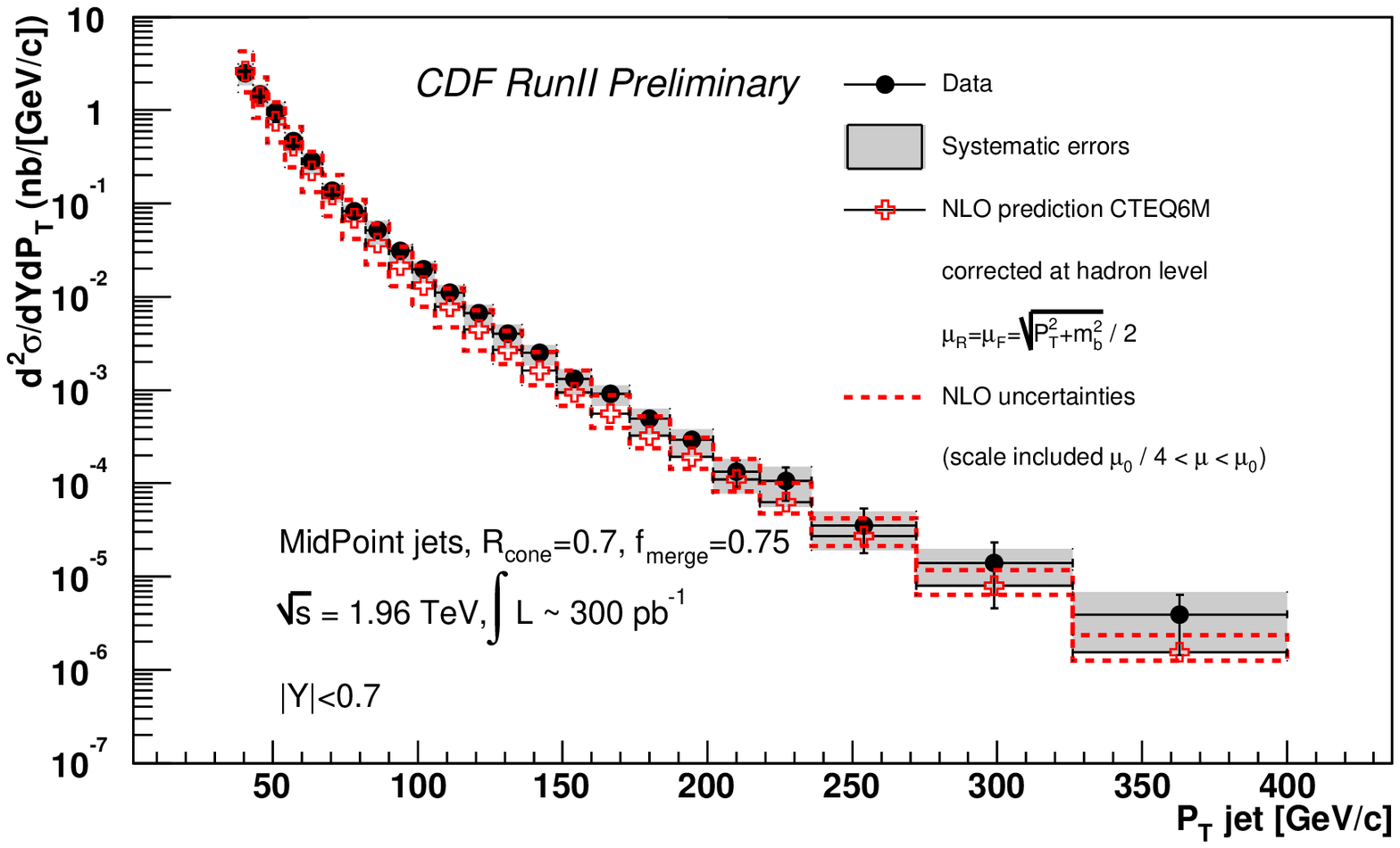}
\includegraphics[height=.20\textheight,width=.31\textheight]{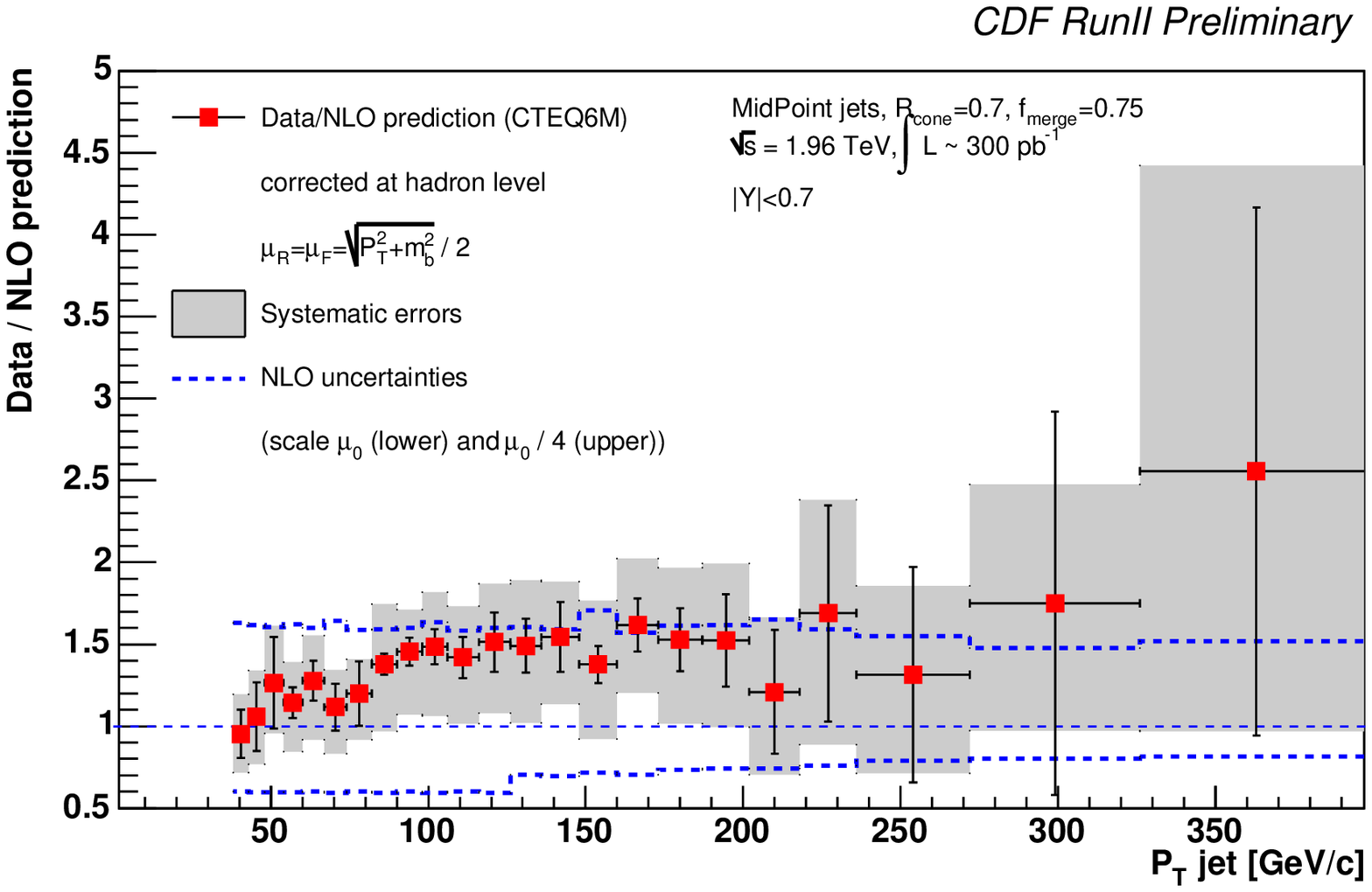}
\caption{Left: inclusive b-jet cross section. Right ratio of measured and NLO cross section.
}
\label{Fig:b}
\end{center}
\end{figure}
CDF has also measured  the inclusive b-jet cross section using $\sim300pb^{-1}$ of data.
Jets are reconstructed using the MidPoint algorithm with a cone radius of 0.7.
Only jets in the central rapidity region ($|y|\le 0.7$) are considered.
The analysis exploits the good tracking capabilities of the detector and rely on b-jet identification
done by secondary vertex reconstruction. The b-tagging algorithm uses displaced tracks
within the jet cone ($R=0.4$).
To determine the heavy flavor content of a tagged jet, thus to extract the fraction of b-jets,
the shape of the secondary vertex mass (SVM) distribution is used.
The presence of neutral particles and neutrinos originating from the secondary vertex
make not possible a full reconstruction of the hadron invariant mass. This results in a dependence on the
jet transverse momentum. The SVM fit is thus performed considering independently each $P^{jet}_T$ bin.
In this way the dependence on $P^{jet}_T$ of the b-fraction and of the tagging efficiency is also
taken into account.
Figure \ref{Fig:b} shows the inclusive b-jet cross section over a $P^{jet}_T$ range between
38 and 400 GeV/c. The main contribution to the systematic error is coming from the jet energy scale.
The measured cross section is compare to NLO pQCD \cite{ref:mf} showing a data to NLO ratio of $\sim1.4$.
The theoretical uncertainty is dominated by the renormalization and factorization
scale choice showing that higher order contributions could play a major role.
D\O~ has also performed an inclusive jet cross section for $\mu-$tagged jets using $\sim300pb^{-1}$.
Jet containing a muon are expected to have an enhanced heavy flavor content. Jets are determined with
the MidPoint algorithm using a cone radius of 0.5 within $|y|<0.5$. The presented results is restricted to
only jets with muons from heavy flavor decay (i.e. muons for whom their creation was within few cm of the
primary vertex). The cross section is corrected to remove muons from pion and kaon decay, and an
unsmearing algorithm is applied. This results in a particle level measurement of $\mu-$tagged jet cross
section that has been found in good agreement with PYTHIA prediction. 
\section*{Acknowledgments}
I would like to acknowledge the EU "Marie Curie" programme for the financial support.

\end{document}